\documentclass[superscriptaddress,aps,prl,twocolumn]{revtex4}
\usepackage{bm}
\usepackage{ulem}
\usepackage{epsfig}
\usepackage{graphicx}
\usepackage{amssymb,amsmath,amsbsy,amsgen,amsfonts}    
\usepackage{dcolumn}
\usepackage{amsthm}
\usepackage{mathrsfs}
\usepackage{latexsym}
\usepackage{array}
\usepackage{color}
\usepackage{amstext}
\allowdisplaybreaks[1]
\usepackage{txfonts}
\usepackage{pifont}

\usepackage{epstopdf} %for texshop on mac

\newcommand{\bra}[1]{\left\langle{#1}\right\vert}
\newcommand{\ket}[1]{\left\vert{#1}\right\rangle}

\newcommand{\be}{\begin{equation}}
\newcommand{\ee}{\end{equation}}
\newcommand{\ba}{\begin{array}}
\newcommand{\ea}{\end{array}}
\newcommand{\bqa}{\begin{eqnarray}}
\newcommand{\eqa}{\end{eqnarray}}

\DeclareSymbolFont{symbols}{OMS}{cmsy}{m}{n}

\begin{document}

\title{Experimental verification of entanglement generated in a plasmonic system}

\author{F. Dieleman}
\affiliation{Experimental Solid State Group, Imperial College London, Blackett Laboratory, SW7 2AZ London, UK}
\author{M. S. Tame}
\email{markstame@gmail.com}
\affiliation{School of Chemistry and Physics, University of KwaZulu-Natal, Durban 4001, South Africa}
\author{Y. Sonnefraud}
\affiliation{Universite Grenoble Alpes, Inst. NEEL, F-38000 Grenoble, France}
\affiliation{2 CNRS, Inst. NEEL, F-38042 Grenoble, France}
\author{M. S. Kim}
\affiliation{Quantum Optics and Laser Science Group, Imperial College London, Blackett Laboratory, SW7 2AZ London, UK}
\author{S. A. Maier}
\email{s.maier@imperial.ac.uk}
\affiliation{Experimental Solid State Group, Imperial College London, Blackett Laboratory, SW7 2AZ London, UK}

\date{\today}

\begin{abstract}
A core process in many quantum tasks is the generation of entanglement. It is being actively studied in a variety of physical settings -- from simple bipartite systems to complex multipartite systems. In this work we experimentally study the generation of bipartite entanglement in a nanophotonic system. Entanglement is generated via the quantum interference of two surface plasmon polaritons in a beamsplitter structure, {\it i.e.}~utilising the Hong-Ou-Mandel (HOM) effect, and its presence is verified using quantum state tomography. The amount of entanglement is quantified by the concurrence and we find values of up to $0.77 \pm 0.04$. Verifying entanglement in the output state from HOM interference is a nontrivial task and cannot be inferred from the visibility alone. The techniques we use to verify entanglement could be applied to other types of photonic system and therefore may be useful for the characterisation of a range of different nanophotonic quantum devices.
\end{abstract}

%\pacs{}

\maketitle

%%%%%%%%%%%%%%%%%%%%%%%%%%%%
%%%%%%%%%%%%%%%%%%%%%%%%%%%%
%%%%%%%%%%%%%%%%%%%%%%%%%%%%
%%%%%%%%%%%%%%%%%%%%%%%%%%%%
{\it Introduction.---} Photonic systems are highly promising candidates for the realisation of quantum technology, such as quantum computing~\cite{Ladd10,KLM01}, quantum communication~\cite{Gis02} and quantum sensing~\cite{Gio11}. At present there is a significant amount of work being undertaken in developing photonics for quantum technology using on-chip waveguides made from a range of dielectric materials~\cite{OBrien05}. While these systems show great promise, they are fundamentally restricted by the diffraction limit~\cite{Tak97}, and cannot be scaled down to dimensions comparable to those used in modern electronic circuitry~\cite{Wal16}. Recent studies have started to explore the potential of using metallic materials, such as gold, graphene and carbon nanotubes, as alternatives for photonic quantum technology~\cite{Tame13}. In these systems, light is coupled to free electrons at a metal surface and propagated along waveguides in the form of a surface plasmon polariton (SPP). The joint light-matter behaviour of SPPs allows the light field to be confined to scales below the diffraction limit, opening up the possibility of highly compact quantum photonic circuitry at the nanoscale. Studies have already shown how one can build single-photon sources~\cite{Akimov07,Kolesov09,Huck11,Cuche11}, switches~\cite{Chang07,Kolchin11,Chang14}, sensing devices~\cite{Fan15,Pooser15,Lee16} and quantum random number generators~\cite{Francis16}, even in the presence of the intrinsic loss that is inherent to compacting light down to the nanoscale~\cite{Tak97}. Experimental work has also shown that SPPs are able to maintain the quantum features of the light used to excite them, including quantum correlations~\cite{DiMartino12} and quantum interference~\cite{Heeres13,Fakonas14,DiMartino14,Cai14,Fujii14}. 

An important feature of quantum plasmonic systems that has not yet been fully explored is their ability to generate entanglement~\cite{Hor09}. Studies have shown that plasmonic systems can be used to preserve entanglement~\cite{Alte02,Fasel2006} and even generate it in a hybrid structure~\cite{Fak15,Wang16}. However, entanglement generated from an all-plasmonic system has not yet been experimentally demonstrated. As entanglement is a resource for many quantum tasks it is vital to understand the entanglement generating capabilities of practical plasmonic systems in order to further develop them as a useful platform for photonic quantum technologies.

In this work, we study the generation of bipartite entanglement in a plasmonic system. In our experiment, two single SPPs are interfered quantum mechanically using a beamsplitter structure via the Hong-Ou-Mandel (HOM) effect~\cite{Hong1987}. This produces a two-plasmon output state that is spatially delocalised over two modes and entangled in the number state degree of freedom~\cite{vanEnk06}: if two SPPs are present in one output mode of the beamsplitter, then none are present in the other and vice versa. The entangled output state can be used as a resource for a number of schemes in quantum information processing~\cite{Ladd10,Gis02,Gio11,KLM01}. We verify the presence of entanglement in the output state using quantum state tomography (QST) and quantify the amount generated. The verification of entanglement produced from the HOM effect is not a straightforward task and cannot be achieved from information about the visibility alone, as visibility does not measure phase information between the output basis states $\ket{0}\ket{2}$ and $\ket{2}\ket{0}$~\cite{Ray11,Marek17}. To the best of our knowledge, there has not yet been any experiment measuring entanglement generated by the HOM effect. On the other hand, the tomography method we use measures phase information between different basis states, as well as information about populations and off-diagonal components of the density matrix~\cite{Schill10,Israel2012}. This provides a full reconstruction of the output state from the plasmonic beamsplitter and a value for the amount of entanglement generated. The techniques used to verify entanglement in our plasmonic system could be easily applied to other quantum photonic systems and may be useful for future characterisation of a range of dielectric and plasmonic devices operating in the quantum regime.
\begin{figure*}[t]
\centering
\includegraphics[width=17cm]{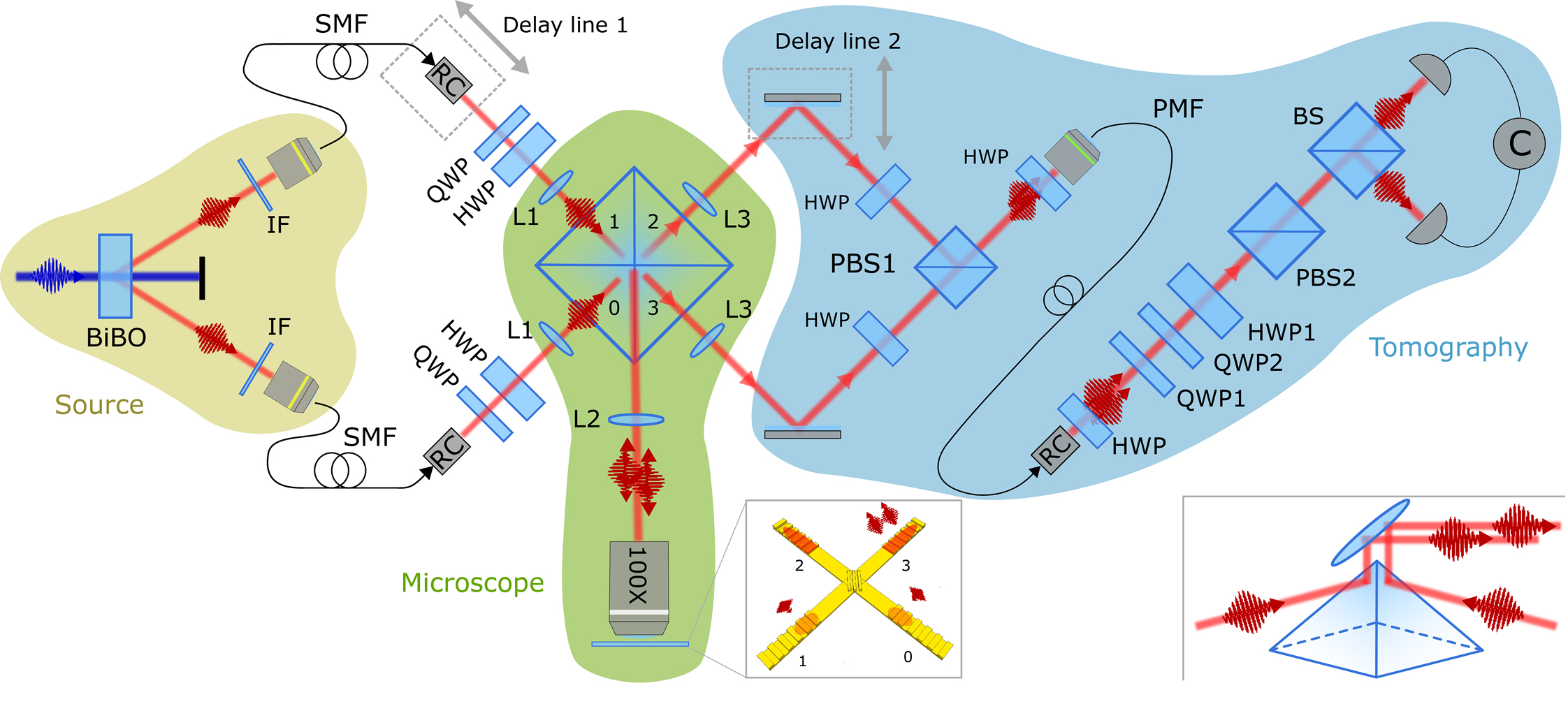}
\caption{Experimental setup for verifying the presence of entanglement generated in a plasmonic system. There are three main stages: The first is the source stage (shown in yellow), where type-I spontaneous parametric down-conversion in a BiBO nonlinear crystal is used to generate pairs of single photons that are coupled into separate single-mode fibers (SMF). These are then sent to the second stage, a microscope stage (shown in green), which consists of an infinity corrected microscope objective with tube lens L2 ($f = 300$~mm). This part of the setup images the plasmonic system on a silver pyramid and enables efficient coupling of the single photons into a plasmonic beamsplitter structure via input gratings (see inset). Lenses L1 ($f = 500$~mm) focus the incoming beams on the image, while lenses L3 ($f=1000$~mm) collimate the beams from the output gratings of the beamsplitter. This pyramid-based configuration allows access to the input and output gratings separately. The output state of the beamsplitter is sent to a tomography stage (shown in blue), where spatial superimposing of the two outputs is achieved using half-wave plates (HWP) and a polarizing beamsplitter (PBS1). This converts the spatial degree of freedom to the polarization degree of freedom. The output of PBS1 is then guided to waveplates via a polarization maintaining fiber (PMF) with a HWP before and after. The wave plates are placed in motorized rotators and integrated with correlation counting software for performing polarization-based coincidence counting using PBS2 and a beamsplitter (BS). This enables the reconstruction of the density matrix of the output state of the plasmonic system.}
\label{fig1} 
\end{figure*}

%%%%%%%%%%%%%%%%%%%%%%%%%%%%
%%%%%%%%%%%%%%%%%%%%%%%%%%%%
%%%%%%%%%%%%%%%%%%%%%%%%%%%%
%%%%%%%%%%%%%%%%%%%%%%%%%%%%
{\it Experimental setup.---} The setup used to verify the generation of entanglement in a plasmonic system is shown in Fig.~\ref{fig1}. Here, two photons are generated in pairs by type-I spontaneous parametric down-conversion (SPDC)~\cite{Hong1986,Grynberg2010}. A birefringent non-linear BiBO crystal is pumped by a 175~mW ($\lambda=404$~nm) diode laser, with bi-photons centred around 808~nm generated stochastically in two modes at half-angles of $3^\circ$ (shown in yellow). Interference filters (IF) centred at 808~nm, with 20~nm FWHM bandwidth, are placed in both modes. The photons are separately guided to a microscope stage (shown in green) by single-mode fibers (SMF) and enter the plasmonic beamsplitter after being collimated by reflective collimators (RC). Polarization rotations are also performed using a quarter-wave plate (QWP) and a half-wave plate (HWP) on each mode to ensure optimal coupling to the input gratings of the beamsplitter (see inset) using two faces of a silver pyramid and a 100$\times$ microscope objective. The output modes of the beamsplitter are then collected using the other two faces of the pyramid (see far right inset) and combined at a tomography stage (shown in blue) using HWPs and a polarizing beamsplitter (PBS), which convert the spatial mode degree of freedom into the polarization degree of freedom. The photons are then sent via a polarization maintaining fiber (PMF) to a section where polarization-sensitive coincidence measurements are performed using single-photon detectors. These enable the reconstruction of the density matrix of the state output by the plasmonic beamsplitter and a quantification of the amount of entanglement generated.

\begin{figure}[t]
\centering
\includegraphics[width=8cm]{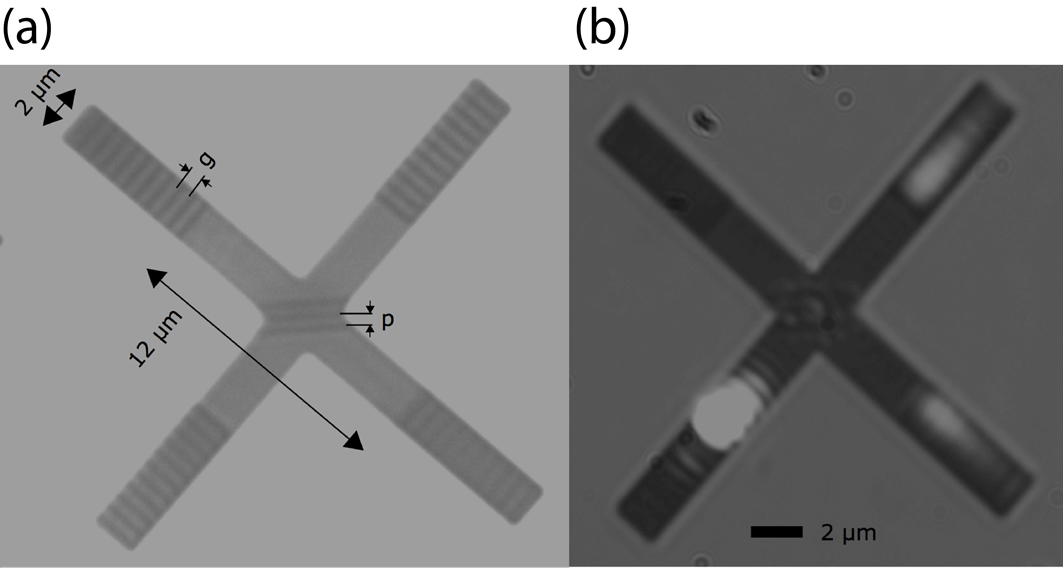}
\caption{Bright-field optical images of the plasmonic beamsplitter taken with the 100$\times$ objective shown in Fig.~\ref{fig1}. {\bf (a)} The input and output gratings consist of 11 rectangular ridges 90~nm high and 290~nm wide, with a grating period $g$ of 580~nm. The Bragg reflector where the waveguides cross consists of three 90~nm high and 150~nm wide ridges, spaced with a period $p$ of 500~nm. {\bf (b)} The beamsplitter structure illuminated with TM-polarized 808~nm laser light.}
\label{BeamSplitter} 
\end{figure}

An optical image of the plasmonic beamsplitter is shown in Fig.~\ref{BeamSplitter}~(a). It consists of two crossing 70 nm thick gold stripe waveguides on a glass substrate, made with electron beam lithography and resistive evaporation of gold~\cite{DiMartino14}. For a free-space wavelength of $\lambda_0 = 808$~nm, the 2 $\mu$m wide stripes have a single leaky SPP mode at the gold-air interface~\cite{Zia2005a, Zia2005b}. At the crossing point of the waveguides a semi-reflective Bragg scatterer consisting of 3 rectangular ridges is patterned. The parameters of these ridges were tuned to provide a close to 50/50 splitting at around $\lambda_0 = 808$~nm. The ends of the waveguides contain ridge gratings which allow the efficient coupling of photons to SPPs and vice versa~\cite{Barnes1996}. Fig.~\ref{BeamSplitter}~(b) shows coupling to SPPs using a laser with wavelength 808~nm. The light is clearly scattering out at both output gratings, showing reliable 50/50 splitting. Further details about the plasmonic beamsplitter and its characterization can be found in Ref.~\cite{DiMartino14}. 

The evolution of the quantum state of the plasmonic beamsplitter can be obtained by taking into account the transformation process. With complex amplitude reflection and transmission coefficients $r$ and $t$, the transformations between annihilation operators for the input and output modes shown in Fig.~\ref{fig1} are $\hat{a}_{2} = t\hat{a}_{0} + r\hat{a}_{1}$ and $\hat{a}_{3} = r\hat{a}_{0} + t\hat{a}_{1}$~\cite{Grynberg2010}. Applying these transformations to the input state $\left|1\right\rangle _0 \left| 1 \right\rangle _1$ gives
\begin{align}
	\begin{split}
	&-\sqrt{2} r^* t^* \left| 2 \right\rangle_2 \left|0 \right\rangle_3   - \sqrt{2} r^* t^* \left| 0 \right\rangle _2 \left|2 \right\rangle _3 \\
	& \hspace{2.5 cm}+ r^{* ^{2}}  \left| 1 \right\rangle _2 \left|1 \right\rangle _3 + t^{* ^{2}}   \left| 1 \right\rangle _2 \left|1 \right\rangle _3.
 	\end{split}\label{BSOut}
\end{align}
The probability of measuring a coincidence, {\it i.e.} one photon at each output, $p$, is proportional to the expectation value $\left\langle \hat{n}_2 \hat{n}_3 \right\rangle$, with $\hat{n}_i$ the photon-number operator $\hat{n}_i = \hat{a}^{\dagger}_i \hat{a}_i$ for mode $i$. In the case of perfect indistinguishability between the two SPPs, the final two terms are combined (interfere coherently) to give $p=|r^{* ^{2}}+t^{* ^{2}}|^2=\left|r\right|^4 + \left|t\right|^4 + 2 Re[r^{* ^{2}} t^2]$, while for perfect distinguishability the final two terms cannot be combined (no interference) and we have $p=\left|r\right|^4 + \left|t\right|^4$. In an experiment with a total number of trials, $N_t$, where two photons are sent in, the number of coincidences measured will be $N=pN_t$. The quality of the interference is then typically expressed by the visibility, the normalized difference between the interference case and the no-interference case: $V = (N _{no-int} - N _{int})/N _{no-int}$. For a symmetric lossless beamsplitter we have $r = i/\sqrt{2}$ and $t = 1/\sqrt{2}$, which gives $N _{int}=pN_t=0$, leading to a visibility of 1. The maximum visibility possible with classical light is 0.5~\cite{Rarity2005}, and a value higher than this directly shows the quantized nature of the electromagnetic field and the bosonic behavior of single excitations -- properties first observed by Hong, Ou and Mandel (HOM)~\cite{Hong1987,Ghosh1987,Ou1989}. Often ignored in dielectric optics, a lossy beamsplitter can have a phase shift between $r$ and $t$ that deviates from $\frac{\pi}{2}$~\cite{Barnett1998}, resulting in a less than perfect visibility, or even an interference peak when $2 Re[r^{* ^{2}} t^2] > 0$~\cite{Uppu2016,Vest2016}. For our plasmonic beamsplitter we first measured values $\left|r\right|^2 = 0.51 \pm 0.05$ and $\left|t\right|^2 = 0.49 \pm 0.05$ using intensity measurements, where the coefficients have been renormalized to remove the contribution to loss. This corresponds to renormalizing the state in Eq.~(\ref{BSOut}) in the case that all measurements are performed in the coincidence basis, as in our experiment. The remaining influence of loss then comes from the relative phase difference between $r$ and $t$. 

To measure the relative phase difference, we placed the plasmonic beamsplitter as the second beamsplitter in a Mach-Zehnder interferometer (MZI), with a lossless beamsplitter as the first beamsplitter. The intensity at the outputs of the MZI are proportional to the total reflection and transmission amplitudes for the system, $|R_{MZ}|^2$ and $|T_{MZ}|^2$. Here, $R_{MZ}=r_p r e^{i\phi_{p2}}+t_p t e^{i\phi_{p1}}$ and $T_{MZ}=r_p t e^{i\phi_{p2}}+r t_p e^{i\phi_{p1}}$~\cite{Loudon2000}, with $r_p=i/\sqrt{2}$ and $t_p=1/\sqrt{2}$ for a symmetric lossless beamsplitter, and $r=|r|e^{i \phi}$ and $t=|t|$ for the plasmonic beamsplitter. The phases $\phi_{p1}$ and $\phi_{p2}$ are propagation phases in each arm of the MZI, and we set $\phi_{p1}=0$ and varied $\phi_{p2}$ relative to it using a delay line. The only remaining unknown parameter is then the phase $\phi$ and comparing the oscillations of output intensities from the MZI as $\phi_{p2}$ changes allows it to be extracted. We find $\phi = 1.21 \pm 0.01$. This value, together with the measured values of $|r|$ and $|t|$ gives a maximum possible visibility of $0.74 \pm 0.01$ for the plasmonic beamsplitter. 

\begin{figure}[b]
\centering
\includegraphics[width=8.5cm]{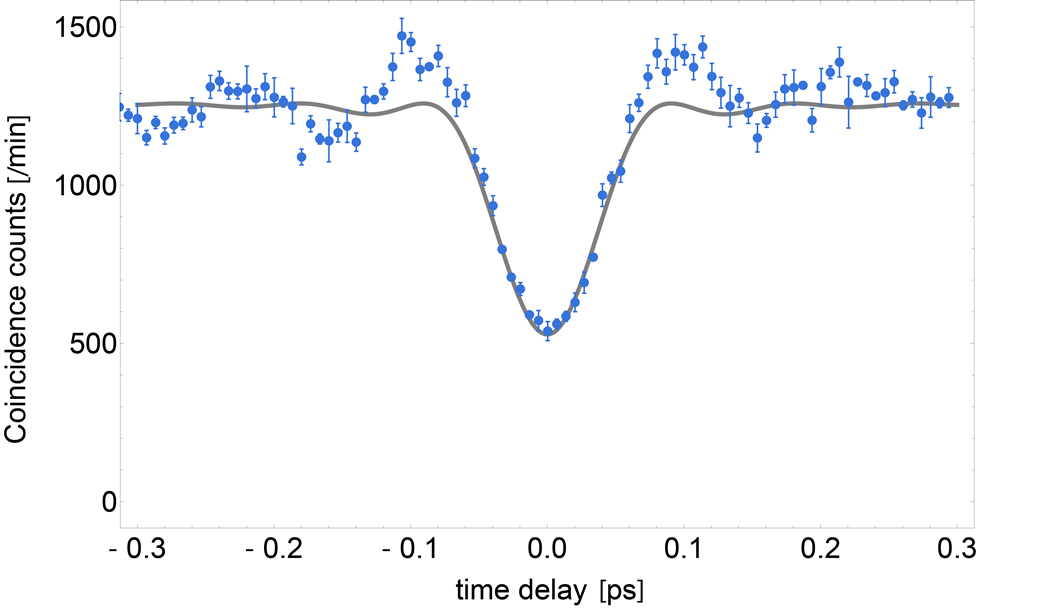}
\caption{HOM interference with the plasmonic beamsplitter. Coincidence counts change as the time delay is varied in delay line 1. Quantum interference is seen at zero time delay, where the coincidences drop to below 50\% of the baseline value (far from zero delay).}
\label{figHOM} 
\end{figure}

In Fig.~\ref{figHOM} we show the dependence of the number of coincidences measured for the plasmonic beamsplitter as a function of the time delay on one of the input photons (delay line 1 in Fig.~\ref{fig1}). Here, the outputs from the plasmonic beamsplitter are sent directly to single-photon detectors. At zero time delay, the photons are indistinguishable in their arrival time at the beamsplitter. As the time delay is increased, the photons can be gradually distinguished based on their arrival time. A visibility of $0.58 \pm 0.01$ is obtained from the figure, clearly above the classical limit of 0.5. The difference between the experimental and theoretical values is likely due to non-ideal overlap of the input and output modes at the Bragg grating in the beamsplitter, which introduces a small amount of distinguishability. Another factor is the inaccuracy in the determination of $\phi$ due to dispersion in the setup~\cite{Mazzotta2016}. This dispersion also creates higher overshoots at the sides of zero delay than expected from the first order model~\cite{DiMartino14}, as shown in Fig.~\ref{figHOM}. Although providing evidence for the quantized nature of SPPs and their bosonic behavior, the HOM effect does not directly imply the generation of entanglement in the plasmonic beamsplitter. For example, coincidence measurements are not able to discriminate between a potential dephased output state $\frac{1}{2}(\left|20\right\rangle \left\langle 20 \right|+ \left|02\right\rangle \left\langle 02 \right|)$ from a given beamsplitter, which has no entanglement, and the entangled state $\frac{1}{\sqrt{2}} (\ket{2}\ket{0}+\ket{0}\ket{2})$. 

\begin{table}[b]
\centering
\begin{center}
\noindent\begin{tabular*}{\columnwidth}{@{\extracolsep{\stretch{1}}}*{4}{r}@{}}
  \hline
  Angle set & $\alpha_{QWP1}$ & $\alpha_{QWP2}$ & $\alpha_{HWP1}$ \\
  \hline
  \hline
$\alpha_1$ ~~~~& 0.102 & 0.440 & 1.740 \\
$\alpha_2$ ~~~~& 1.803 & 1.144 & -0.330 \\
$\alpha_3$ ~~~~& 2.010 & 1.083 & -0.464 \\
$\alpha_4$ ~~~~& 0.102 & -0.236 & 1.402 \\
$\alpha_5$ ~~~~& 0.232 & -0.427 & 1.241 \\
$\alpha_6$ ~~~~& 0.439 & -0.488 & 1.107 \\
$\alpha_7$ ~~~~& $\pi/4$ & $\pi/4$ & $13\pi/16$ \\
$\alpha_8$ ~~~~& $\pi/4$ & $\pi/4$ & $7\pi/8$ \\
$\alpha_9$ ~~~~& $\pi/4$ & $\pi/4$ & $15\pi/16$ \\
\hline            
\end{tabular*}
  \end{center}    
\caption{Wave plate angles used in the tomography stage. All angles are given in radians and are between the fast axes of the plates and the vertical axis.}
\label{tab:angles} 
\end{table}

In order to verify the presence of entanglement in the output state of the plasmonic beamsplitter the spatial mode degree of freedom of the output photons is transferred to the polarization degree of freedom and quantum state tomography (QST) is then performed in the polarization basis. The HWPs and PBS1 after the plasmonic beamsplitter in Fig.~\ref{fig1} apply the following transformations to the annihilation operators: $\hat{a}_2 \to \hat{a}_V$ and $\hat{a}_3 \to \hat{a}_H$, where both polarizations are now in the same spatial mode. For example, the output state from ideal HOM interference would be transferred to the state $\frac{1}{\sqrt{2}} (\ket{2}_H\ket{0}_V+e^{i \phi}\ket{0}_H\ket{2}_V)$ in the output mode of PBS1. Here, a possible phase $\phi$ has been included due to delay line 2, which is used to ensure the temporal overlap of the two modes at PBS1 so that coherence is maintained. In order to allow for full generality of the output state, we consider single-mode multiphoton polarization-entangled states with a definite photon number $N$ given by
\begin{align}
	\ket{\psi_N}=\sum _{n=0} ^{N} c_n \left| n \right\rangle _H \left| N - n \right\rangle _V, 
	\label{multiphoton}
\end{align}
with $\sum _{n=0} ^{N} \left|c_n\right|^2 = 1$. Here, we set $N=2$, as we are considering pairs of single photons produced at the source and measurements performed in the coincidence basis at the detectors. The state $\ket{\psi_2}$ and mixtures of it can be fully determined by a QST method consisting of three wave plates and a polarizer, shown as QWP1, QWP2, HWP1 and PBS2 in the tomography stage of Fig.~\ref{fig1}. The HWP-PMF-HWP chain preceding the waveplates is only to transfer the state from PBS1 to another part of the setup where stability of components can be monitored. QST is then performed by measuring second-order intensities, where any second-order intensity measurement can be related to the second-order coherences between the two polarizations of the mode exiting PBS1~\cite{Schill10}. The second-order intensities are measured using a 50/50 beamsplitter and coincidence measurements. Schilling~\textit{et al.}~\cite{Schill10} have shown that it is possible to calculate the 9 unique second-order coherences $\left\langle (\hat{a}^{\dagger}_H) ^{2-w}  (\hat{a}^{\dagger}_V) ^{w} \hat{a}_H ^{2-y}  \hat{a}_V ^{y} \right\rangle$ (with $ w,y = \{0,1,2\}$), by making 9 measurements of the second-order intensity with different angle combinations $\alpha = (\alpha _{QWP1}; \alpha _{QWP2}; \alpha _{HWP})$ of the wave plates. The set of 9 different angle combinations $\alpha _i$ have to be chosen in such a way that the 9 equations relating the intensities to the coherences are independent, allowing one to solve for the set of 9 coherences. The angles used are given in Tab.~\ref{tab:angles}.

The measured coherences are then directly related to elements of the density matrix by the relation $\rho _{2-y,2-w} = \left\langle (\hat{a}^{\dagger}_H) ^{2-w}  (\hat{a}^{\dagger}_V) ^{w} \hat{a}_H ^{2-y}  \hat{a}_V ^{y} \right\rangle / \sqrt{(2-y)!y!(2-w)!w!}$~\cite{densityelements}. From Eq.~(\ref{multiphoton}), these elements enter the density matrix as $\rho=\sum_{i,j=0}^{2} \rho_{i,j} \ket{i}_{~H}\bra{j} \otimes \ket{2-i}_{~V} \bra{2-j}$. Due to the conversion from path degree of freedom to polarization degree of freedom in the tomography stage at PBS1, the density matrix in the two spatial modes output from the plasmonic beamsplitter is then $\rho=\sum_{i,j=0}^2 \rho_{i,j} \ket{i}_{~1}\bra{j} \otimes \ket{2-i}_{~2} \bra{2-j}$. To simplify notation we have used the mode labelling 1 and 2, previously modes 3 and 2 respectively in Fig.~\ref{fig1}. Solving the set of equations relating the intensities to coherences directly to obtain the density matrix elements can deliver nonphysical results due to experimental noise (the matrix has to be positive semidefinite, Hermitian and have a trace of 1)~\cite{Israel2012}. A maximum likelihood approach is therefore used to find the closest fit of a general positive semidefinite matrix to the data~\cite{James2001}. This is then the most likely physical state from which the data could have been obtained.
 \begin{figure}[t]
\centering
\includegraphics[width=8cm]{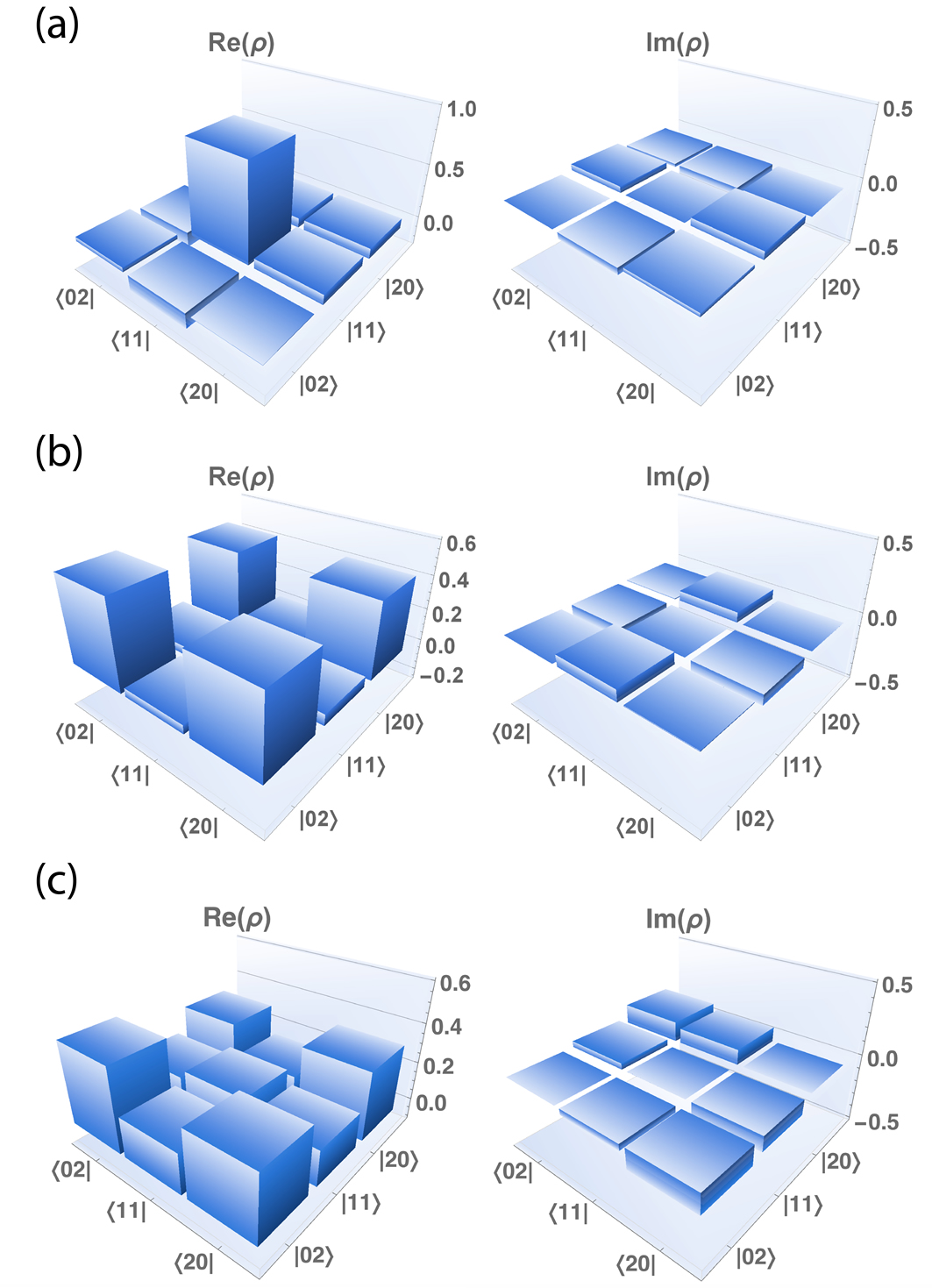}
\caption{Experimental density matrices for the different states used in the investigation. \textbf{(a)} Input photon state $\left|11\right\rangle$. \textbf{(b)} Output state from photonic HOM. \textbf{(c)} Output state from plasmonic HOM.}
\label{Tomography} 
\end{figure}

%%%%%%%%%%%%%%%%%%%%%%%%%%%%
%%%%%%%%%%%%%%%%%%%%%%%%%%%%
%%%%%%%%%%%%%%%%%%%%%%%%%%%%
%%%%%%%%%%%%%%%%%%%%%%%%%%%%

{\it Results.---} Before performing QST on the HOM output state, the density matrix of the $\left| 1 \right\rangle \left| 1 \right\rangle$ input state was reconstructed. For this, the setup is used without the microscope stage and the photons coming from the reflective collimators are directly superimposed on PBS1. The result is shown in Fig.~\ref{Tomography}~(a). The fidelity $ F \left( \rho , \sigma \right) \equiv Tr \left[ \sqrt{\rho ^{1/2} \sigma \rho ^{1/2}} \right] ^2 $ allows a comparison of the measured state $\rho$ to an expected state $\sigma$, with a value of 0 and 1 meaning no and a full overlap between $\rho$ and $\sigma$, respectively. The measured state shows a fidelity of $0.93 \pm 0.01$ for $\sigma = \ket{\psi}\bra{\psi}$ (with $\ket{\psi}=\ket{1}\ket{1}$), showing both the high quality of the generated bi-photon state and the reliability of the technique. The tomography was then performed on the HOM output state from a photonic beamsplitter and a plasmonic beamsplitter at the centre of the dip~\cite{delayline1}, with the results shown in Fig.~\ref{Tomography}~(b) and (c). For the photonic case the whole microscope stage was replaced by a cube beamsplitter, creating a test of the scheme for a near-ideal case due to the high visibility, $V=0.93\pm 0.01$, and high count rates of the photonic HOM. A fidelity of $0.92 \pm 0.01$ was obtained when comparing the output state with the ideal two-photon HOM state, showing a good overlap. The diagonal elements (populations) of the density matrix shown in  Fig.~\ref{Tomography}~(b) are 0.52, 0.03 and 0.45 for $\left|02 \right\rangle$,  $\left|11 \right\rangle$ and  $\left|20 \right\rangle$ respectively. When there is no interference between the photons one would expect a population distribution of 0.25, 0.5 and 0.25. The value of 0.03 obtained thus implies a visibility of the HOM interference of 0.94, in good agreement with the value obtained by directly measuring the photonic HOM dip. 

On the other hand, the density matrix for the plasmonic case shown in Fig.~\ref{Tomography}~(c) has populations of 0.42, 0.24 and 0.34. The higher $\left| 1 1 \right\rangle$ component is again consistent with the visibility obtained by directly measuring the HOM interference. The fidelity of the plasmonic output state compared with the ideal case is $F= 0.64 \pm 0.07$. As can be seen in Fig.~\ref{Tomography}~(c), the presence of off-diagonal components in the imaginary part of the density matrix for the plasmonic case signify a non-zero phase in the output state. By maximising the fidelity of the state compared with the ideal HOM state with an arbitrary phase we find $\phi \simeq -0.4$. This phase is set by delay line 2 and can be adjusted close to zero in principle, however its value fluctuates over time. We only collect data over a stable period for a fixed phase and the data shown is for one such instance. The large error in the fidelity for the plasmonic output state compared to the photonic one is due to reduced counts over the stability period.

We now quantify the entanglement in the photonic and plasmonic output states. To do this we use the concurrence, a measure of entanglement for two-qubit states~\cite{Wootters1998}. To transform the basis $\{\left|0\right\rangle \left|2\right\rangle,\left|1\right\rangle \left|1\right\rangle,\left|2\right\rangle \left|0\right\rangle \}$ to a system of a two-qubit state, a few basic transformations are performed. First, the Hilbert space is expanded to include the $\left|2\right\rangle \left|2\right\rangle$ state. Elements in the density matrix corresponding to this state are set to zero as the probability of four photons being produced by the source is negligible compared to two photons at the pump power used in our experiment~\cite{DiMartino12}. Second, the state $\left|0\right\rangle \left|0\right\rangle$ is included in the Hilbert space. As we are measuring in the coincidence basis, the measured state has no contribution from this state and elements in the density matrix corresponding to it are also set to zero. It is worth noting that due to loss, most of the time both photons will be absorbed in the plasmonic beamsplitter and the state generated will be $\left|0\right\rangle \left|0\right\rangle$. However, by measuring in the coincidence basis we are postselecting out these events and measuring states generated by the beamsplitter when both photons from the source make it to the detectors. 

Finally, a filtering of the $\left|1 \right\rangle \left|1\right\rangle$ state is performed. The filter checks if there is exactly one photon in a mode or not, and if so the state is rejected, otherwise it is accepted. As the filter does not require any knowledge of the number of photons in the other mode it is a local filtering operation and cannot increase the amount of entanglement in the state~\cite{Ray11}. The filtering is performed by setting the density matrix elements with a $\left|1 \right\rangle \left|1\right\rangle$ component equal to zero. With these transformations, both modes are now represented in the $\{\left|0\right\rangle, \left| 2 \right\rangle \}$ basis (each a qubit), resulting in a $4 \times 4$ density matrix, $\rho_t$. We then calculate the concurrence of the transformed and filtered state, $\rho_t$, to obtain a lower bound on the entanglement of the original physical state, $\rho$, output from the beamsplitter. The filtered concurrence is given by $C(\rho_t) = max \{0, \lambda_1 - \lambda_2 - \lambda_3 - \lambda_4 \}$, with $\lambda _i$ the eigenvalues in decreasing order of the matrix $\sqrt{\sqrt{\rho_t} \tilde{\rho_t} \sqrt{\rho_t}}$ and $\tilde{\rho_t}$ is the density matrix of the spin-flipped state of $\rho_t$~\cite{Wootters1998}. The filtered concurrence ranges from 0 to 1, with 0 signifying no entanglement and 1 corresponding to full entanglement. To take into account the filtering operation and the fact that the original state from the beamsplitter had $\left|1\right\rangle \left|1\right\rangle$ components, we multiply the filtered concurrence by the fraction of the original state in the $\left|2\right\rangle\left|0\right\rangle$ and $\left|0\right\rangle\left|2\right\rangle$ subspace: $P = \rho _{20} + \rho_{02}$~\cite{Ray11}. The lower bound for the entanglement of the original state $\rho$ when there is no filtering is thus $C_{nf}=P\,C(\rho_t)$. For the photonic HOM interference a value of $C_{nf}=0.92 \pm 0.01$ was obtained and for the plasmonic case we find $0.62 \pm 0.12$ for the state shown in Fig.~\ref{Tomography}~(c). Values of $C_{nf}$ up to $0.77 \pm 0.04$ were observed for plasmonic output states that had different values of the phase $\phi$. Both the photonic and plasmonic cases show an entangled state is created in the interference at the respective beamsplitters and highlight that not only is it possible to preserve entanglement in a plasmonic system, but also to generate it, despite the presence of loss.

%%%%%%%%%%%%%%%%%%%%%%%%%%%%
%%%%%%%%%%%%%%%%%%%%%%%%%%%%
%%%%%%%%%%%%%%%%%%%%%%%%%%%%
%%%%%%%%%%%%%%%%%%%%%%%%%%%%

{\it Discussion.---} In this work we have shown it is possible to generate entanglement in a completely plasmonic structure. The density matrix in the coincidence basis of the HOM output state was reconstructed with a polarization-based quantum state tomography scheme and a measure of entanglement, the concurrence, was used to quantify the entanglement in the output state.  Values of up to 0.77 $\pm$ 0.04 were obtained. To our knowledge this is the first time the generation of entanglement in an all-plasmonic structure has been directly probed. The results add further strength to the view that plasmonic structures can be used in experiments and applications in quantum optics and quantum information~\cite{Tame13}. Interestingly, we also saw that the loss in the system opened up a new parameter regime not fully explored in dielectric optics -- it was found that the phase shift in a plasmonic beamsplitter can strongly deviate from the lossless case, potentially enabling new effects like two-photon absorption in a linear medium~\cite{Barnett1998, Lee2012} and the generation of related entangled output states. We expect our work to be useful for state characterization in such schemes in the future.

%%%%%%%%%%%%%%%%%%%%%%%%%%%%
%%%%%%%%%%%%%%%%%%%%%%%%%%%%
%%%%%%%%%%%%%%%%%%%%%%%%%%%%
%%%%%%%%%%%%%%%%%%%%%%%%%%%%
{\it Acknowledgments.---} We thank Will McCutcheon for helpful discussions. This research was supported by the Marie Sklodowska-Curie Early Stage Researcher programme, the Marie Curie Training Network on Frontiers in Quantum Technologies (FP7/2007-2013), the European Office of Aerospace Science and Technology EOARD, the South African National Research Foundation, the University of KwaZulu-Natal Nanotechnology Platform and the South African National Institute for Theoretical Physics. M. S. Kim acknowledges the Royal Society and the UK EPSRC (FP/K034480/1). S. A. Maier acknowledges the Lee-Lucas Chair in Physics.

%%%%%%%%%%%%%%%%%%%%%%%%%%%%
%%%%%%%%%%%%%%%%%%%%%%%%%%%%
%%%%%%%%%%%%%%%%%%%%%%%%%%%%
%%%%%%%%%%%%%%%%%%%%%%%%%%%%


\begin{thebibliography}{99}
\bibitem{densityelements} This can be calculated in general by writing the expectation values for two modes as $\left\langle (\hat{a}^{\dagger}_1) ^{N-w}  (\hat{a}^{\dagger}_2) ^{w} \hat{a}_1 ^{N-y}  \hat{a}_2 ^{y} \right\rangle ={\rm Tr}(\rho (\hat{a}^{\dagger}_1) ^{N-w}  (\hat{a}^{\dagger}_2) ^{w} \hat{a}_1 ^{N-y}  \hat{a}_2 ^{y})=\sum_{i,j=0}^N\rho_{ij} \phantom{.}_1\bra{j} \phantom{.}_2\bra{N-j}(\hat{a}^{\dagger}_1) ^{N-w}  (\hat{a}^{\dagger}_2) ^{w} \hat{a}_1 ^{N-y}  \hat{a}_2 ^{y} \ket{i}_1 \ket{N-i}_2$. Evaluating this expression gives $ \left\langle (\hat{a}^{\dagger}_1) ^{N-w}  (\hat{a}^{\dagger}_2) ^{w} \hat{a}_1 ^{N-y}  \hat{a}_2 ^{y} \right\rangle=\rho _{N-y,N-w}\sqrt{(N-y)!y!(N-w)!w!}$. In the main text we have used the mode labelling $1=H$ and $2=V$.
\end{thebibliography}
\end{document}